\documentclass[twocolumn,aps,prl] {revtex4} 
\usepackage{graphicx}
\begin{document}
\pagestyle{empty} 
\title{Rolling friction for hard cylinder and sphere on viscoelastic solid}
\author{  
B.N.J. Persson} 
\affiliation{IFF, FZ-J\"ulich, D-52428 J\"ulich, Germany}
\affiliation{School of Chemistry, Tel Aviv University, Tel Aviv, 69978, Israel}

\begin{abstract}
We calculate the friction force acting on a hard cylinder or spherical ball rolling on a flat
surface of a viscoelastic solid. The rolling friction coefficient depends 
non-linearly on the normal load and the rolling velocity. 
For a cylinder rolling on a viscoelastic solid characterized by a single relaxation time
Hunter has obtained an exact result for the rolling friction, and our 
result is in very good agreement with his result for this limiting case. 
The theoretical results are also
in good agreement with experiments of Greenwood and Tabor. We suggest that measurements
of rolling friction over a wide range of rolling velocities and temperatures
may constitute an useful way to determine the viscoelastic modulus of rubber-like materials.
\end{abstract}
\maketitle


{\bf 1 Introduction}

Rubber friction is a topic of huge practical importance, e.g., for tires, rubber seals,
wiper blades, conveyor belts and syringes
\cite{Book,Grosch,Persson1,JPCM,P3,theory1,theory2,theory3,theory4,theory5,theory6,wear,Mofidi,Creton}. 
Many experiments have been performed with a hard spherical ball 
rolling on a flat rubber substrate\cite{Tabor1,Tabor2,Varadi1}. 
Nearly the same friction force is observed during
sliding as during rolling, assuming that the interface is lubricated and that the sliding velocity 
and fluid viscosity are such that a thin lubrication film is formed with a thickness much
smaller than the indentation depth of the ball, but larger 
than the amplitude of the roughness on the surfaces\cite{Tabor1}.
The results of rolling friction experiments have often been analyzed using 
a very simple model of Greenwood and Tabor\cite{Tabor1}, which however
contains a (unknown) factor $\alpha$, which represent the fraction of the input elastic energy lost as a result of the internal friction
of the rubber. In this paper we present a very simple theory for the friction force acting on a hard cylinder or spherical ball
rolling on a flat rubber surface. 
For a cylinder rolling on a viscoelastic solid characterized by a single relaxation time
Hunter\cite{Hunter} has obtained an exact result for the rolling friction, and our 
result is in very good agreement with his result for this limiting case. 

\vskip 0.3cm
{\bf 2 Theory}

Using the theory of elasticity (assuming an
isotropic viscoelastic medium), one can calculate the displacement field $u_i$
on the surface $z=0$ in response to the surface stress distributions
$\sigma_i= \sigma_{3i}$. 
Let us define the Fourier transform
$$u_i({\bf q},\omega) = {1\over (2\pi)^3} \int d^2x \  dt \ u_i({\bf x},t) {\rm e}^{-i({\bf q}
\cdot {\bf x}-\omega t)}$$
and similar for $\sigma_i({\bf q},\omega )$. Here ${\bf x}=(x,y)$ and ${\bf q}=(q_x,q_y)$
are two-dimensional vectors. In Ref. \cite{Persson1} we have shown that 
$${u}_i ({\bf q}, \omega ) = 
M_{ij} ({\bf q}, \omega ) \sigma_j ({\bf q},\omega )$$
or, in matrix form,
$${\bf u} 
({\bf q}, \omega ) = M ({\bf q}, \omega ) {\bf \sigma} ({\bf q},\omega )$$
where the matrix $M$ is given in Ref. \cite{Persson1}.

We now assume that $\mid \nabla u_z({\bf x})\mid < 1$ and that
the surface stress only acts in the $z$-direction so that
$$ u_z ({\bf q},\omega ) = 
M_{zz} ({\bf q}, \omega ) {\bf \sigma}_z({\bf q},\omega ).\eqno(1)$$
Since in the present case $\omega$ is of order $v q$ (where $v$ is the sliding or rolling velocity)
we get $\omega/c_T q = v/c_T << 1$ in most cases
of practical interest, where $c_{\rm T}$ is the transverse sound velocity in the rubber. 
In this case (see Ref. \cite{Persson1}):
$$\left (M_{zz}\right )^{-1} =- {E q \over 2(1-\nu^2)}.\eqno(2)$$
It is interesting to note that if, instead of assuming that the surface stress
act in the $z$-direction, we assume that the displacement
${\bf u}$ point along the $z$-direction, then
$$\sigma_z({\bf q},\omega) = \left ( M^{-1}\right )_{zz}({\bf q}, \omega) u_z({\bf q},\omega)$$
where in the limit $\omega/c_Tq <<1$,
$$\left (M^{-1}\right )_{zz} 
=- {2E q (1-\nu) \over (1+\nu)(3-4\nu)}$$
which differ from (2) only with respect to a factor $4(1-\nu)^2/(3-4\nu)$. For rubber-like
materials ($\nu \approx 0.5$) this factor is of order unity. Hence, practically identical results
are obtained independently of whether one assumes that the interfacial
stress or displacement vector is perpendicular to the nominal contact surface. In reality, neither of
these two assumptions hold strictly, but the 
result above indicate that the theory is 
not sensitive to this approximation.

Now, assume that
$$\sigma_z({\bf x}, t) = \sigma_z ({\bf x}-{\bf v}t)$$
then
$$\sigma_z ({\bf q},\omega ) = {1\over (2 \pi)^3} \int d^2x dt \ \sigma_z ({\bf x}- {\bf v}t)
 \ {\rm e}^{-i({\bf q}\cdot
{\bf x}-\omega t)}$$
$$=\delta (\omega - {\bf q}\cdot {\bf v})  
 \ \sigma_z ({\bf q})\eqno(3)$$
where 
$$\sigma_z ({\bf q}) = {1\over (2 \pi)^2} \int d^2x \ \sigma_z ({\bf x})
{\rm e}^{-i{\bf q}\cdot
{\bf x}}$$
If $F_{\rm f}$ denote the friction force then the energy dissipated during the
time period $t_0$ equals 
$$\Delta E = F_{\rm f}  v t_0.\eqno(4)$$ 
But this energy can also be written as
$$\Delta E = \int d^2xdt \ \dot {\bf u}\cdot \sigma$$
$$= \left (2\pi \right )^3 \int d^2q d\omega \ (-i\omega){\bf u} ({\bf q},\omega)\cdot
{\bf \sigma} (-{\bf q},-\omega )\eqno(5)$$
where $\omega = {\bf v}\cdot {\bf q}$.
Substituting (1) in (5) and using (3) and that
$$[\delta (\omega - {\bf q}\cdot {\bf v})]^2 = (t_0/2\pi )  
 \ \delta (\omega - {\bf q}\cdot {\bf v}),$$ 
gives
$$\Delta E = \left (2 \pi \right )^2 t_0 
\int d^2q \ (-i\omega)$$
$$\times M_{zz}(-{\bf q}, -\omega) \sigma_z({\bf q}) \sigma_z(-{\bf q}).$$
Comparing this expression with (4) gives the friction force
$$F_{\rm f} = {\left (2\pi \right )^2\over v} 
\int d^2q \ (-i\omega) \ M_{zz} (-{\bf q}, -\omega) 
\sigma_z({\bf q}) \sigma_z(-{\bf q}).$$
Since $F_{\rm f}$ is real and since $\sigma_z(-{\bf q}) = \sigma^*_z({\bf q})$ we get
$$F_{\rm f} = {\left (2\pi \right )^2\over v} 
\int d^2q \ \omega {\rm Im} M_{zz} (-{\bf q}, -\omega) 
|\sigma_z({\bf q})|^2 .\eqno(6a)$$
Using (2) we can also write 
$$F_{\rm f} = {2 \left (2\pi \right )^2\over v} 
\int d^2q \ {\omega \over q} {\rm Im} {1\over E_{\rm eff} (\omega)} 
|\sigma_z({\bf q})|^2.\eqno(6b)$$
where $E_{\rm eff} = E/(1-\nu^2)$. 
In principle, $\nu$ depends on frequency but the factor $1/(1-\nu^2)$
varies from $4/3\approx 1.33$ for $\nu = 0.5$ (rubbery region) 
to $\approx 1.19$ for $\nu = 0.4$ (glassy region) and we can neglect the weak
dependence on frequency. 
Within the assumptions given above, equation (6) is exact. Note that
even if we use $\sigma_z({\bf q})$ calculated to zero order in ${\rm  tan} \delta$, the friction force
(6) will be correct to linear order in ${\rm tan}\delta$. 
We also note that the present approach is very general and flexible. For example, instead of a semi-infinite
solid as assumed above one may be interested in a thin viscoelastic film on a hard flat substrate. In this case,
(assuming slip-boundary conditions) the
$M({\bf q},\omega)$-function which enter in (6a) is determined by\cite{theory4,Carbon}
$$M^{-1}=-{Eq \over 2(1-\nu^2)}S$$
$$S={(3-4\nu){\rm cosh}(2qd)+2(qd)^2-4\nu(3-2v)+5\over 
(3-4\nu){\rm sinh}(2qd)-2qd}$$
Note that as $d\rightarrow \infty$, $S\rightarrow 1$ and the present result reduces to (2). 
Substituting this in (6a) gives the rolling
friction for a sphere (or cylinder) on a thin rubber film adsorbed to a hard flat substrate.
We now apply the theory to (a) a rigid cylinder and (b) 
a rigid sphere rolling on a semi-infinite viscoelastic solid.

\vskip 0.3cm

{\bf 2.1 Cylinder}

Consider a hard cylinder (radius $R$ and length $L_y>>R$) rolling on a viscoelastic solid.
The same result is obtained during sliding if one assume lubricated contact and if one can neglect
the viscous energy dissipation in the lubrication film. As discussed above, when calculating the friction force
to linear order in ${\rm tan}\delta$ we can neglect dissipation when calculating the contact pressure
$\sigma_z(x)$ and assume that the stress is of the Hertz form. 
Thus if we introduce a coordinate system with the $y$-axis parallel to the cylinder axis and with the origin of the $x$-axis
in the middle of the contact area (of width $2a$), then the contact stress for $-a <x <a$: 
$$\sigma_z(x) = {2 f_{\rm N}\over  \pi a}  \left [1-\left ({x\over a}\right )^2\right ]^{1/2}\eqno(7)$$ 
where $f_{\rm N} = F_{\rm N}/L_y$ the load per unit length of the cylinder.
The half-width of the contact area in the Hertz contact theory is 
$$a = \left ({4 f_{\rm N} R \over \pi E_{\rm eff}} \right )^{1/2}\eqno(8)$$
where we take $E_{\rm eff}$ to be $|E_{\rm eff}(\omega)|$ with $\omega= {\bf q}\cdot {\bf v}$. 
Now, note that
$$\int d^2x \left [1-\left ({x\over a}\right )^2\right ]^{1/2} e^{-i{\bf q} \cdot {\bf x}}$$
$$=2\pi\delta(q_y)\int_{-a}^a dx \ \left [1-\left ({x\over a}\right )^2 \right ]^{1/2} e^{-iq_x x}$$
$$=2\pi  \delta(q_y) a \int_{-1}^1 dz \ \left [1-z^2 \right ]^{1/2} e^{-iq_x a z}$$
$$=2\pi \delta(q_y) {\pi \over q_x} J_1(aq_x)\eqno(9)$$
Using (7) and (9) gives
$$\sigma({\bf q}) = {f_{\rm N}\over \pi a q_x } \delta(q_y)  J_1(q_xa)\eqno(10)$$
Note that this expression satisfies
$$(2\pi )^2 \sigma ({\bf q=0})= 
\int d^2x \ \sigma_z({\bf x}) = F_{\rm N}$$
where we have used that $\delta (q_y=0) = L_y/2\pi$.
Substituting (10) in (6b) dividing by $F_{\rm N}$, and using that
$$[\delta(q_y)]^2 = (L_y/2\pi) \delta(q_y),$$
gives the friction coefficient:
$$\mu =  {8 f_{\rm N}\over \pi} \int_0^\infty dq_x \ {\rm Im} {1\over E_{\rm eff} (q_xv)} 
{1\over (a q_x)^2} J_1^2(q_xa)\eqno(11)$$

\begin{figure}
\includegraphics[width=0.45\textwidth,angle=0]{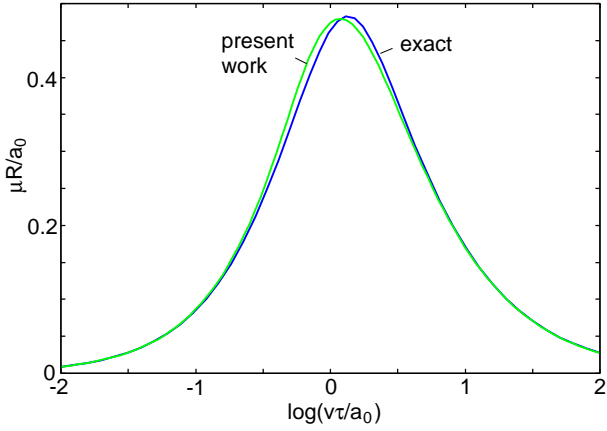}
\caption{\label{compare}
The friction coefficient (times the radius $R$ of the cylinder and divided by the
half-width $a_0$ of the static contact region) as a function of $v\tau /a_0$, where $v$ is the rolling velocity, $\tau$ the
rubber viscoelastic relaxation time (for $E_1/E_0 = 10$).
We compare the exact result (blue curve) of Hunter\cite{Hunter} with 
the prediction of Eq. (11) (green curve).}
\end{figure}

Let us now assume the simplest possible viscoelastic modulus characterized by a single relaxation time $\tau$:
$$E_{\rm eff} (\omega) = {E_1(1-i\omega \tau)\over E_1/E_0 - i\omega \tau }\eqno(12)$$
where $E_1/E_0$ is the ratio between the high frequency and low frequency 
modulus. 
In Fig. \ref{compare} we show the
friction coefficient (times the radius $R$ of the cylinder and divided by the
half-width $a_0=a(v=0)$ of the static contact region) as a function of $v\tau /a_0$, where $v$ is the rolling velocity.
We have assumed $E_1/E_0 = 10$. 
We compare the exact result (blue curve) of Hunter\cite{Hunter} (the same result was obtained by Goryacheva\cite{Gor}) with 
the prediction of Eq. (11) (green curve). Note that some distance away from the maximum the agreement between the
two curve is perfect. This is expected because these regions correspond to small ${\rm tan}\delta$ where 
Eq. (11) should be essentially exact. Close to the maximum a small difference occur between the two curves, but from
a practical point of view this is not important, since real rubber exhibit some non-linearity
making any linear viscoelasticity theory only approximately valid anyhow.

At this point we empathize that (6) is basically exact, and if one could calculate the contact pressure $\sigma_z (x)$ 
[or rather the Fourier Transform $\sigma_z ( q_x)$]
exactly, then (6) should give the exact result, e.g., the result of Hunter for the cylinder case, 
and assuming the simple viscoelastic modulus (12). Note that the
contact pressure will not be symmetric around the midpoint when the dissipation in the rubber is included in the analysis.
Still, to linear order in ${\rm tan}\delta$ one can neglect this effect 
[since (6) is explicitly already linear in ${\rm tan}\delta$],  and the analysis 
above shows that this is a remarkable accurate approximation, which is an
interesting result in its own right, and makes it possible to apply the theory to a wide set of problems.

\vskip 0.3cm

{\bf 2.2 Sphere}

Consider a hard spherical (radius $R$) ball rolling on a viscoelastic solid.
As discussed above, when calculating the friction force
to linear order in ${\rm tan}\delta$ we can neglect dissipation when calculating the contact pressure $\sigma_z({\bf x})$
and assume that the stress is of the Hertz form:
$$\sigma_z ({\bf x}) = {3 F_{\rm N}\over 2 \pi r_c^2}  \left [1-\left ({r\over r_c}\right )^2\right ]^{1/2}\eqno(13)$$ 
where $r=|{\bf x}|$ is the distance from the center of the contact area and $F_{\rm N}$ the load.
The radius of the contact area in the Hertz contact theory is 
$$r_c = \left ({3 F_{\rm N} R \over 4 E_{\rm eff}} \right )^{1/3}\eqno(14)$$
where we take $E_{\rm eff}$ to be $|E_{\rm eff}(\omega)|$ with $\omega= {\bf q}\cdot {\bf v}$. 
Now, note that
$$\int d^2x \left [1-\left ({r\over r_c}\right )^2\right ]^{1/2} e^{-i{\bf q} \cdot {\bf x}}$$
$$=\int_0^{r_c} dr r \left [1-\left ({r\over r_c}\right )^2\right ]^{1/2} \int_0^{2\pi} d\phi \ e^{-iqr{\rm cos} \phi}$$
$$= 2 \pi \int_0^{r_c} dr r \left [1-\left ({r\over r_c}\right )^2\right ]^{1/2} J_0(qr)$$ 
$$= 2\pi r_c^2 \int_0^1 dz z \left [1-z^2 \right]^{1/2} J_0(qr_cz)$$
$$= 2\pi r_c^2 (qr_c)^{-3}\left [{\rm sin} (qr_c)- qr_c {\rm cos} (qr_c) \right ]\eqno(15)$$ 
Using (13) and (15) gives
$$\sigma({\bf q}) = {3 F_{\rm N}\over 4 \pi^2} {1\over (q r_c)^3} 
\left [{\rm sin} (qr_c)- qr_c {\rm cos} (qr_c) \right ]\eqno(16)$$
Note that this expression satisfies
$$(2\pi )^2 \sigma ({\bf q=0})= 
\int d^2x \ \sigma_z({\bf x}) = F_{\rm N}$$
Substituting (16) in (6b) and dividing by $F_{\rm N}$ gives the friction coefficient:
$$\mu = {9 F_{\rm N}\over 2 \pi^2} \int_0^\infty dq q \int_0^{2\pi} d\phi \ 
{\rm cos}\phi \ {\rm Im} {1\over E_{\rm eff} (qv{\rm cos}\phi)}$$
$$ \times
{1\over (qr_c)^6}\left [{\rm sin} (qr_c)-qr_c {\rm cos}(qr_c)\right ]^2\eqno(17)$$
Eq. (17) is very general with no restriction on the viscoelastic properties of the rubber or on the sliding velocity $v$
(assuming that $v$ is small enough that the effect of frictional heating is negligible, and assuming $v << c_{\rm T}$).

\begin{figure}
\includegraphics[width=0.45\textwidth,angle=0]{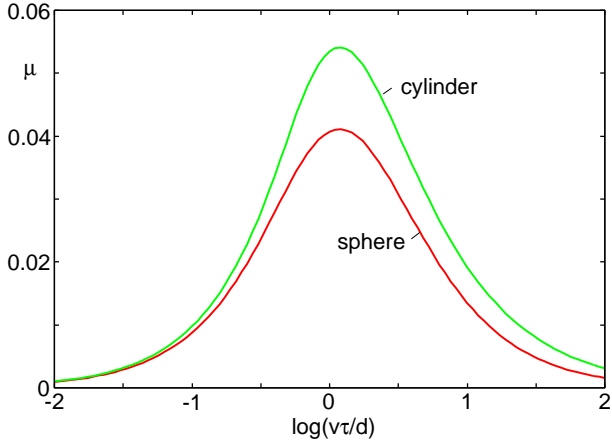}
\caption{\label{mu}
The friction coefficient as a function of $v\tau /d$, where $v$ is the rolling velocity, $\tau$ the
rubber viscoelastic relaxation time, and $d=a_0$ for the cylinder (upper curve) or $d=r_0$ for the sphere (lower curve),
rolling on a rubber substrate described by a simple
viscoelastic model (see Eq. (12)). Here $a_0$ is half the width of the cylinder-substrate contact area and
$r_0$ the radius of the contact region for the sphere, both in the limit of vanishing rolling velocity.
In the calculation the radius of the sphere and of the cylinder are both $R=1 \ {\rm cm}$ and the load was
chosen so that the average contact pressure for the stationary contact was the same (equal to 
about $44 \ {\rm kPa}$). For $E_0=1 \ {\rm MPa}$ and $E_1/E_0 = 10$.}
\end{figure}

In Fig. \ref{mu} we show the rolling friction coefficient 
as a function of $v \tau /r_0$ for a sphere, and as a function of $v \tau/a_0$ for a cylinder. 
The ball has the radius $R=1 \ {\rm cm}$ and is
squeezed against a rubber surface with the load $F_N = 0.15 \ {\rm N}$ giving a static contact area
with the radius $r_0 = r_c(v=0)= 0.10 \ {\rm cm}$. The average contact pressure $p_{\rm a} = F_N/(\pi r_0^2) = 44.3 \ {\rm kPa}$
is the same as for the cylinder case shown in the same figure.
The cylinder has the radius $R=1 \ {\rm cm}$ and is
squeezed against a rubber surface with the load $f_N = 100 \ {\rm N/m}$ giving a static contact area
with the width $a_0= 0.11 \ {\rm cm}$. The average contact pressure is $p_{\rm a} = f_N/(2a_0) = 44.3 \ {\rm kPa}$.
In the calculation we have used the viscoelastic modulus $E_{\rm eff} (\omega)$ given in Eq. (12).
Note that at very low sliding velocities where the (average) contact pressure are the same the rolling frictions
are nearly identical.

\begin{figure}
\includegraphics[width=0.45\textwidth,angle=0]{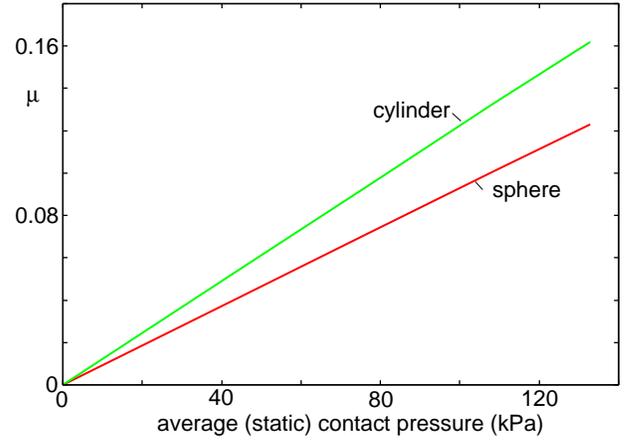}
\caption{\label{pmu}
The maximum (as a function of the velocity $v$) of the rolling friction coefficients for a sphere and a cylinder, as a function of 
the average (static) contact pressure.
For the same parameters as in Fig. \ref{mu}.}
\end{figure}

In figure \ref{pmu} we show the 
maximum (as a function of the velocity $v$) of the rolling friction coefficients for a sphere and a cylinder, as a function of 
the average pressure in the static contact area. Note that $\mu$ varies nearly linearly with the (static) contact pressure, and using
(8) and (14) it follows that the rolling friction coefficient $\mu \sim f_{\rm N}^{1/2}$ for the cylinder 
and $\mu \sim F_{\rm N}^{1/3}$ for the sphere.

\begin{figure}
\includegraphics[width=0.45\textwidth,angle=0]{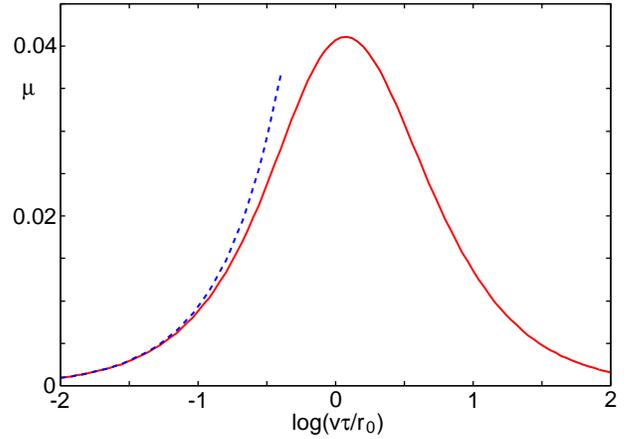}
\caption{\label{asym}
The friction coefficient as a function of $v\tau /r_0$
for the sphere.
The full line is using the full theory [Eq. (17)] and the dashed line using the asymptotic, low velocity,
result (19).
For the same parameters as in Fig. \ref{mu}.}
\end{figure}

Let us now consider the limiting case when the rolling velocity $v$ is so small that only the low-frequency
viscoelastic modulus is relevant. We also assume that that $E(\omega)$ is given by (12).
If $\omega \tau << 1$ for all relevant frequencies, we get ${\rm Re} E(\omega) \approx E(0) = E_0$ 
and ${\rm Im} E(\omega) \approx -\omega \tau E_0 (1-E_0/E_1)$.
We also get $|E(\omega)| \approx E_0$. 
Substituting these results in (17) gives after some simplifications
$$\mu \approx  {9 I \over 2} {p_{\rm a}\over E_0}
\left (1-{E_0\over E_1}\right ) {v\tau \over r_0} \eqno(18)$$
where $p_{\rm a} = F_{\rm N}/(\pi r_0^2)$ is the (average) static contact pressure, and where
$$I= \int_0^\infty d\xi \ \xi^{-4} \left ( {\rm sin} \xi- \xi {\rm cos} \xi \right )^2 \approx 0.52$$
Note that $\mu \sim p_{\rm a}$ with is consistent with Fig. \ref{pmu}. 
If we define the frequency $\omega = v/r_0$ and note that in the present case ${\rm tan} \delta = (1-E_0/E_1) \omega \tau$ we can write
$$\mu \approx {9 I\over 2} {p_{\rm a} \over E_0} {\rm tan}\delta . $$
Since $ 9I/2 \approx 2.34$ we get
$$\mu \approx 2.34 {p_{\rm a} \over E_0} {\rm tan}\delta . \eqno(19)$$
In Fig. \ref{asym} we compare this asymptotic (low velocity) result with the full theory [Eq. (17)].
For the limiting case studied above Greenwood and Tabor have shown that
$$\mu = {9\pi \over 64} {p_{\rm a}\over E_0} \alpha, \eqno(20)$$
where $\alpha$ is the fraction of the input elastic energy lost as a result of the internal damping of the rubber.
Thus we can write
$$\alpha \approx 5.3 {\rm tan}\delta. \eqno(21)$$
Greenwood and Tabor analyzed rolling (and lubricated sliding) friction data using (20). They obtained the best fit to the experimental
data by using an $\alpha$ which was almost a factor of two larger than 
obtained from cyclic (low frequency) simple uniaxial loading-unloading measurements of rubber strips, where the energy loss due to
hysteresis was about 0.35 of the maximal elastic energy of deformation. For a linear viscoelastic solid the latter is given by
$\pi {\rm tan}\delta$ (see Appendix A). Our theory predict that $\alpha$ (see (21)) is about $1.7$ times larger than $\pi {\rm tan}\delta$, and using
this result gives very good agreement with the measured data presented in Ref. \cite{Tabor1}. 
The reason for why rolling friction experiments result in an $\alpha$-parameter which is larger than expected from
uni-axial tension tests is related to the very different nature of the time-dependent deformations: a rubber volume element
below a rolling sphere (or cylinder) undergoes, for some fraction of the interaction time, time-dependent deformations where the total elastic
energy is nearly constant while the stress directions are changing and energy is being lost (see discussion in Ref. \cite{Tabor1,Green}). 

\vskip 0.3cm
{\bf 3 Rolling friction on rubber}

\begin{figure}
\includegraphics[width=0.45\textwidth,angle=0]{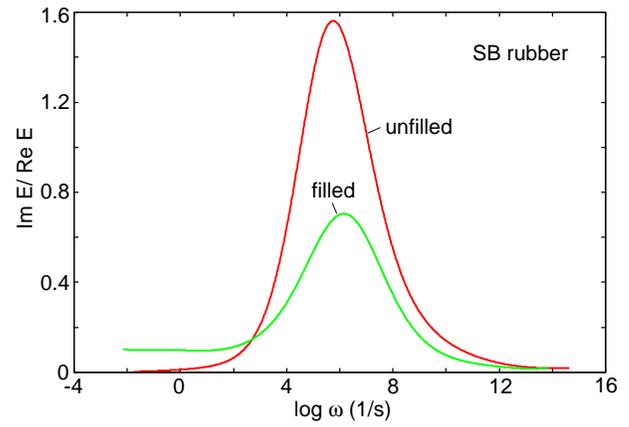}
\caption{\label{SB.Manfred.tan.delta.filled.unfilled}
The loss tangent ${\rm tan}\delta = {\rm Im}E(\omega)/{\rm Re}E(\omega)$ at $T=20 \ {\rm C}$, as a function
of frequency $\omega$ for an unfilled and filled SB rubber.
}
\end{figure}

We have calculated the rolling friction when a hard cylinder with radius $R=1 \ {\rm cm}$ is rolling
on a Styrene-Butadiene (SB) copolymer rubber surface at room temperature. The viscoelastic modulus of the rubber
has been measured, and in Fig. \ref{SB.Manfred.tan.delta.filled.unfilled} we show
the loss tangent ${\rm tan}\delta = {\rm Im}E(\omega)/{\rm Re}E(\omega)$ at $T=20 \ {\rm C}$, as a function
of frequency $\omega$ for an unfilled and filled SB-rubber. Note that ${\rm tan}\delta$ for the filled SB-rubber has a tail 
extending to very low frequencies. This correspond to relaxation processes in the rubber with very long relaxation times
(see Fig. B.2 in \cite{JPCM}).
This effect result from the increase in the activation barrier for polymer rearrangement processes for rubber molecules 
bound to (or close to) filler particles (in this case carbon particles).  

\begin{figure}
\includegraphics[width=0.45\textwidth,angle=0]{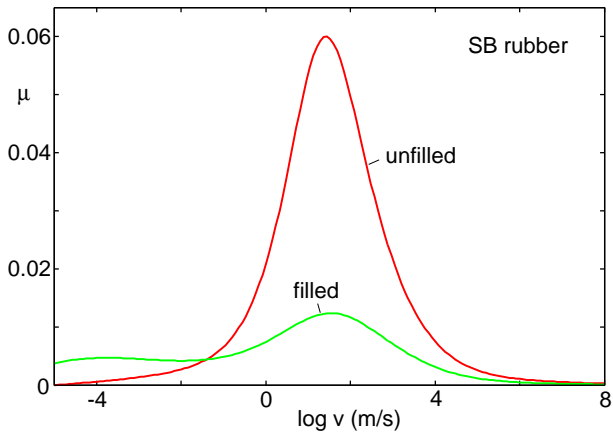}
\caption{\label{logv.muRolling.Manfred.filled.unfilled.SB}
The rolling friction coefficient for a hard cylinder with radius $R= 1 {\rm cm}$ rolling
on unfilled and filled SB rubber at $T=20 \ {\rm C}$. The normal force on the
cylinder per unit length is $10 \ {\rm N/cm}$.  
}
\end{figure}

The rolling friction coefficient is shown in Fig. \ref{logv.muRolling.Manfred.filled.unfilled.SB}
when the normal force per unit length is $10 \ {\rm N/cm}$. The rolling friction exhibit a very strong temperature
dependence given by the WLF shift factor $a_T$: reducing the temperature by $10 \ {\rm C}$  result in 
a shift of the rolling friction curve towards lower velocities by approximately one decade. 
At low frequencies (or low rolling velocities) the unfilled SB rubber is elastically much softer than the filled rubber.
Thus, the half width $a_0$ of the static Hertz contact region is $1.4 \ {\rm mm}$ for the unfilled rubber
but only $0.6 \ {\rm mm}$ for the filled rubber. At high rolling velocities these values becomes much smaller owing to the stiffening of
the effective elastic modulus at high frequencies. Note also that the rolling friction 
for filled SB rubber has a tail extending towards lower rolling velocities,
which has the same origin as the tail in ${\rm tan}\delta$ towards lower frequencies, 
involving rubber molecules bound to the filler particles.

\vskip 0.3cm
{\bf 4 Discussion}

Consider a rolling cylinder (or ball) in a reference frame where the center-of-mass velocity $v$ vanish.
The rolling friction gives a moment $F_{\rm f} R$ at the center of the cylinder. Since the shear stress at the
interface is assumed to vanish, this moment must arise from a contact pressure $\sigma_z(x)$ which is asymmetric around
the center line of the cylinder. That is,
$$F_{\rm f} R = L_y \int dx \ x \sigma_z(x).$$
Thus the calculation of the rolling resistance gives information about how the centroid of the 
contact pressure shift away from the symmetry line $x=0$ due to the delayed response of the rubber caused
by the internal friction of the rubber (characterized by the relaxation time $\tau$ in (12)). 
This information could in principle be used to improve the theoretical treatment given above, but since the 
simple theory we use seems to be very accurate, we will not consider this point further here.

The study presented above assumes that the adhesional interaction between the rubber and the hard ball can be neglected.
Most rubber of engineering interest are strongly cross-linked and have fillers making them relative stiff and reducing
the role of adhesion. Adhesion can also be removed (or reduced) by lubricating the interface or by adsorbing small inert solid 
particles on the rubber surface, e.g., talc. In these latter cases, the rolling and sliding friction may be nearly the same
as indeed observed in some experiments\cite{Tabor1}. For very soft rubber and for 
clean surfaces, during rolling or sliding an opening (and a closing) crack
will propagate at the interface, and associated with this may be very strong energy dissipation, which may dominate the rolling
or sliding friction (see, e.g., Ref. \cite{Review}.). 
We note finally that measurements of the rolling friction as a function of velocity $v$ and temperature $T$
may be a very useful way of determining the viscoelastic
modulus $E(\omega)$, see Appendix B. 

\vskip 0.3cm
{\bf 5 Summary and conclusion}

We have presented a very general and flexible approach to calculate the rolling resistance of hard
objects on viscoelastic solids. The theory can be applied to materials with arbitrary (e.g., measured) viscoelastic modulus
$E(\omega)$. The theory can be applied to both spheres and cylinders rolling on semi-infinite viscoelastic solids, or on
a thin viscoelastic film adsorbed on a rigid flat substrate, or even more complex situations for
which the $M({\bf q},\omega)$-function can be calculated.
For a cylinder rolling on a viscoelastic solid characterized by a single relaxation time $\tau$,
Hunter has obtained an exact solution for the rolling resistance. We have shown that for this limiting case our theory gives 
almost the same as the result as obtained by Hunter. We have compared the rolling resistance of a sphere with that of a cylinder 
under similar circumstances. 
Measurements of the rolling friction as a function of velocity and temperature
may be a very useful way of determining the viscoelastic modulus. 
\vskip 0.5cm
{\bf Acknowledgments}

I thank J.A. Greenwood for drawing my attention to the work of S.C. Hunter and I.G. Goryacheva, and for
supplying the numerical data for the blue curve in Fig. 1 (theory of Hunter). I also thank
him and A. Nitzan for useful comments on the text. I thank A. Nitzan for the warm hospitality
during a 2 weeks stay at Tel Aviv University where this work was done. 
I thank M. Kl\"uppel for the (measured) viscoelastic modulus of the filled and unfilled SB-rubber.
This work, as part of the European Science Foundation EUROCORES Program FANAS, was supported from funds 
by the DFG and the EC Sixth Framework Program, under contract N ERAS-CT-2003-980409.

\vskip 0.5cm
{\bf Appendix A}

Consider a strip of rubber in (slow) loading-unloading and assume that the strain
$\varepsilon = \varepsilon_0 {\rm sin}(\omega t)$, $0<t<t_0$ with $\omega t_0 = \pi$.
The energy dissipation (per unit volume)
$$U_{\rm diss} = \int_0^{t_0} dt \ \sigma \dot \varepsilon $$
$$= \int_0^{t_0} dt \ {\varepsilon_0 \over 2i} \left (E(\omega) e^{i\omega t}-E(-\omega)e^{-i\omega t}\right )$$
$$\times {\varepsilon_0 \over 2} \omega\left (e^{i \omega t}+e^{-i\omega t}\right ) 
={1\over 2} \varepsilon_0^2 \omega t_0 {\rm Im}E(\omega)$$
We are considering very low frequencies so that the maximum elastic energy
$$U_{\rm el} \approx {1\over 2} E(0) \varepsilon_0^2\approx {1\over 2} {\rm Re} E(\omega) \varepsilon_0^2$$ 
Thus we get 
$$\alpha = {U_{\rm diss}\over U_{\rm el}} = \pi {{\rm Im} E(\omega) 
\over {\rm Re} E(\omega) } \approx \pi {\rm tan \delta}\eqno(A1)$$

\vskip 0.5cm
{\bf Appendix B}

The viscoelastic modulus $E(\omega )$ is a complex quantity and at first one may think that it impossible
to determine both ${\rm Re}E(\omega)$ and ${\rm Im} E(\omega)$ from a knowledge of a single function $\mu(v)$
which depends on $E(\omega)={\rm Re}E(\omega)+i{\rm Im}E(\omega)$. However, $E(\omega)$ is (assumed to be) 
a causal linear response function so that ${\rm Im} E(\omega)$
can in fact be obtained from ${\rm Re}E(\omega)$ using a Kramer-Kronig relation. Thus we have one known function $\mu (v)$
and only one unknown function, e.g, ${\rm Re}E(\omega)$, and $E(\omega)$ can in principle
be determined uniquely from $\mu(v)$. The best way of doing this is to 
note that $E(\omega )$ can be written as
$$E(\omega) = E_1-\int_0^\infty d\tau {H(\tau) \over 1-i\omega \tau}\eqno(B1)$$
In numerical calculations one may discretize the relation (B1) and it is usually enough to include $n\approx 15$ 
relaxation times $\tau_k$ with $\tau_{k+1}\approx 10\tau_k$. Thus we 
expand $E(\omega)$ on the form
$$E(\omega) = E_1-\sum_{k=1}^n {H_k \over 1 -i\omega \tau_k}\eqno(B2)$$
where the $H_k$ and the relaxation times $\tau_k$ are real positive quantities. 
Next, form the quantity
$$V=\sum_i [\mu_{\rm meas}(v_i) - \mu_{\rm theory}(v_i;H_1, ..., H_n)]^2g_i\eqno(B3)$$
where $g_i$ are suitable chosen weight coefficients (e.g., $g_i=1$) and where $v_i$ are the velocities for which the
rolling friction has been measured. In (B3) we have indicated that the theory expression for the rolling friction 
coefficient depends on the numbers $H_k$. 
The determination of $H_k$ is a problem in multidimensional minimization of $V$, 
and can be performed using different methods, e.g.,
the Monte Carlo method or the Amoeba method, see Ref. \cite{Num}.

\end{document}